\documentclass{emulateapj}
\usepackage{natbib}
\usepackage{times}
\usepackage[colorlinks=true,citecolor=black,filecolor=blue,linkcolor=blue,urlcolor=blue]{hyperref}

\begin{document}

\journalinfo{UCB-NPAT-11-007}
\shorttitle{Metallicity and the Rate of Type Ia Supernovae}
\shortauthors{Kistler et al.}

\title{The Impact of Metallicity on the Rate of Type Ia Supernovae}

\author{Matthew D.~Kistler\altaffilmark{1,2,6},
K.~Z.~Stanek\altaffilmark{3,4},
Christopher S.~Kochanek\altaffilmark{3,4},
Jos{\'e}~L.~Prieto\altaffilmark{5,7}, and
Todd~A.~Thompson\altaffilmark{3,4,8}
}

\affil{$^1$California Institute of Technology, Mail Code 350-17, Pasadena, CA 91125, USA}
\affil{$^2$Department of Physics, University of California, Berkeley and Lawrence Berkeley National Laboratory, Berkeley, CA 94720, USA}
\affil{$^3$Center for Cosmology \& Astro-Particle Physics, The Ohio State University, Columbus, OH 43210, USA}
\affil{$^4$Department of Astronomy, The Ohio State University, Columbus, OH 43210, USA}
\affil{$^5$Carnegie Observatories, 813 Santa Barbara Street, Pasadena, CA 91101, USA}
\altaffiltext{6}{Einstein Fellow}
\altaffiltext{7}{Hubble, Carnegie-Princeton Fellow}
\altaffiltext{8}{Alfred P.~Sloan Fellow}



\begin{abstract}
The metallicity of a star strongly affects both its evolution and the properties of the stellar remnant that results from its demise.  It is generally accepted that stars with initial masses below $\sim 8\,M_\odot$ leave behind white dwarfs and that some sub-population of these lead to Type~Ia supernovae.  However, it is often tacitly assumed that metallicity has no effect on the rate of SNe~Ia.  We propose that a consequence of the effects of metallicity is to significantly increase the SN~Ia rate in lower-metallicity galaxies, in contrast to previous expectations.  This is because lower-metallicity stars leave behind higher-mass white dwarfs, which should be easier to bring to explosion.  We first model SN~Ia rates in relation to galaxy masses and ages alone, finding that the elevation in the rate of SNe~Ia in lower-mass galaxies measured by LOSS is readily explained.  However, we then see that models incorporating this effect of metallicity agree just as well.  Using the same parameters to estimate the cosmic SN~Ia rate, we again find good agreement with data up to $z\approx2$.  We suggest that this degeneracy warrants more detailed examination of host galaxy metallicities.  We discuss additional implications, including for hosts of high-$z$ SNe~Ia, the SN~Ia delay time distribution, super-Chandrasekhar SNe, and cosmology.
\end{abstract}


\section{Introduction}
\label{section:introduction}

The end result of the evolution of stars that produce white dwarfs  is often a Type Ia supernova explosion, through a single-degenerate channel (e.g., \citealt{Whelan73}), double-degenerate channel (e.g., \citealt{Iben:1984iz,Webbink:1984ti}), or both.  Since stellar evolution is obviously affected by metallicity, there is no a priori reason why the rate of SNe~Ia should not significantly depend on metallicity.  From a theoretical standpoint, a preference for high metallicity was proposed by \citet{Kobayashi:1998ii}, whose single-degenerate model required a minimum metallicity of $\sim\! 0.1\,Z_\odot$ in order to produce SNe~Ia.  A similar preference for higher metallicity was seen in the single degenerate models of \citet{Langer:2000fs}.  However, the strong predictions offered by these models, such as no SNe~Ia in dwarf galaxies and the outskirts of spirals, were not confirmed observationally (e.g., \citealt{Prieto}).  The viability of the single-degenerate channel to produce the majority of SNe~Ia has been debated from both observational (e.g., \citealt{Leonard:2007nh,Simon:2009hk,Gilfanov:2010ip}) and theoretical viewpoints \citep{Ruiter:2009dk,Kasen,Hachisu}, as has the double degenerate scenario \citep{Pakmor:2009yx,Pakmor:2012qs,Fryer:2010jx,Shen:2011ey,Dan:2012dt,Zhu:2012xt}.

From the viewpoint of stellar evolution, we expect an opposite sign for the dependence of the rate of Type Ia SNe on metallicity.  Stars of lower metallicity at a given mass generally produce more massive white dwarfs according to stellar evolution calculations (e.g., \citealt{Umeda:1998ij,Marigo:2007xq,Meng:2007ni}), which should make it easier for them to reach the Chandrasekhar mass and explode.  For example, according to \citet{Marigo:2007xq}, a single star with an initial mass of $3\,M_{\odot}$ will leave behind a $\simeq\! 0.7\,M_{\odot}$ white dwarf when evolved at solar metallicity ($Z_\odot \!=\! 0.019$), while a star with the same initial mass at much lower metallicity ($Z \!=\! 0.001$) will leave behind a $>\! 0.8\,M_{\odot}$ white dwarf.  Due to the steepness of the stellar initial mass function (IMF), this leads to a larger number of SN~Ia progenitors.  Obviously, producing a Type~Ia SN explosion is a more complicated process than just evolving single stars (for example, see the discussion of common envelope phase treatment in \citealt{Ruiter:2009dk}).  However, the observed rate of Type Ia SNe implies that a large fraction ($\sim\! 2 \!-\! 40$\%) of all $3 \!\lesssim M \lesssim 8\,M_{\odot}$ stars will explode as one (e.g., \citealt{Maoz:2007xw}), which suggests that the evolution leading to SN~Ia production cannot be ``fragile''.

In this paper, we propose that the Type Ia supernova rate has a strong dependence on stellar metallicity and examine the potential observational signatures in order to test the overall sign of this effect.  We construct a simple model that examines the SN~Ia rate in galaxies, as a function of galaxy mass, age, and metallicity, and in the universe at large.  This is largely motivated by the measurements of the nearby SN~Ia rates reported by the Lick Observatory SN Search (LOSS: \citealt{Leaman,Li:2010ii,Li:2010iii}).  These measurements and their interpretation are discussed in Section~\ref{section:ii}.

We present our model in Section~\ref{section:iii}, accounting first for a dependency of the SN~Ia rate on galaxy mass and age alone.  In Section~\ref{section:iv}, we expand upon our model to incorporate this possible effect of metallicity in regulating the SN~Ia rate and discuss a variety of implications and competing effects, such as the dependence of stellar radius on $Z$.  We extend this in Section~\ref{section:v} into a treatment of the cosmic SN~Ia rate, addressing ``prompt'' and ``delayed'' SNe.  In Section~\ref{section:vi}, we discuss methods to discern the role of metallicity, including the differential examination of SNe~Ia host galaxies, the hosts of high-$z$ SNe, galactic chemical evolution, SNe in galactic halos, and super-Chandrasekhar SNe, and effects on cosmological studies.

\section{The Type Ia Supernova Rate in Galaxies}
\label{section:ii}

At present, the most complete and systematic search for nearby supernovae was conducted over the past decade by the Lick Observatory SN Search, with results recently detailed in \citet{Leaman} and \citet{Li:2010ii,Li:2010iii}.  Here, we briefly discuss the implications of the LOSS findings for our present study.  Of particular interest are the results pertaining to Type Ia supernovae.

\begin{figure}[t!]
\includegraphics[width=3.38in,clip=true]{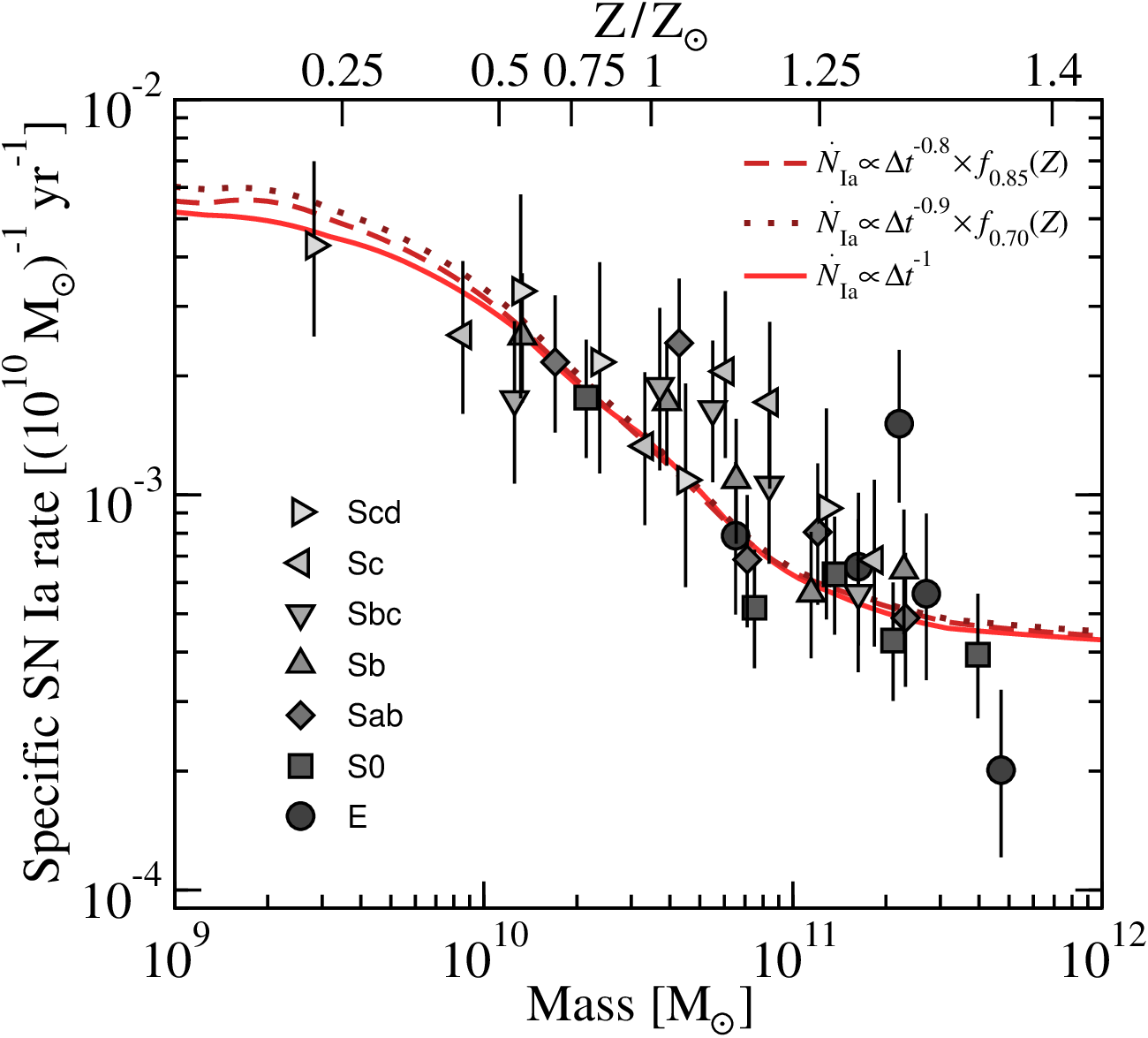}
\caption{The specific rate of Type Ia supernovae versus host galaxy mass.  Shown are data from LOSS for galaxies grouped by Hubble type \citep{Li:2010iii}.  Our models are also displayed, which assume either a $\Delta t^{-1}$ delay time distribution alone ({\it solid line}) or an additional dependence on stellar metallicity ({\it dashed, dotted lines}; see text).\\
\label{Iarate}}
\end{figure}

In Fig.~\ref{Iarate}, we display the specific SN~Ia rate (rate per unit mass) versus galaxy mass as measured by LOSS \citep{Li:2010iii}.  One is first struck by the steep dependence of this specific rate on galaxy mass.  This variation of over an order of magnitude demands a physical explanation.  The cause should be distinct from the origin of a similar pattern seen in the specific core-collapse supernova rate by LOSS, which likely arises mainly from the dependence of the specific star formation rate (SFR) on galaxy mass \citep{Li:2010iii}.

Importantly, we also see that at a fixed mass the SN~Ia rate does not vary greatly between galaxies of different Hubble type.  This suggests that by examining a large set of galaxies one can arrive at the global behavior of SNe Ia.  For our later use, we proceed to translate the LOSS measured specific SN~Ia rates in galaxies of various Hubble types from a function of galactic mass into one of galactic {\it metallicity}.  To do this, we convert between galactic mass and median metallicity using the relation derived from SDSS data in \citet{Gallazzi:2005df}, as shown by the upper axis of Fig.~\ref{Iarate}.  This technique effectively averages over a large representative galaxy population similar to that sampled by LOSS.

\section{A Simple Galactic Rate Model}
\label{section:iii}
We first attempt to explain the rate variations in galaxies of different mass as due to an age effect alone.  Since there is a delay from stellar birth to SN~Ia explosion, a galaxy's SN~Ia rate depends upon the age of its white dwarf population.  This is typically quantified by an empirical or theoretical delay-time distribution (DTD), which results in a SN~Ia rate that can be simply written as
\begin{equation}
	\dot{N}_{\rm Ia}(t) = \int_{t_0}^{t} dt^\prime \, \phi(t-t^\prime) \,  \dot{\rho}_*(t^\prime),
	\label{eq:rate}
\end{equation}
where $t_0$ is the age of the universe when SN~Ia progenitor stars first formed and $\phi(t-t^\prime)$ is the DTD, which maps between the rate of star formation at time $t^\prime$, $\dot{\rho}_*(t^\prime)$, and the SN~Ia rate at a later time $t=t^\prime+\Delta t$.  Eq.~(\ref{eq:rate}) can be used to calculate the expected SN~Ia rate of an individual galaxy or the universe as a whole, given a properly normalized $\dot{\rho}_*(t)$ (for the cosmic SN~Ia rate, we will use the SFR density).  Recent studies have suggested that $\phi$ roughly takes a $\Delta t^{-1}=(t-t^\prime)^{-1}$ form (e.g., \citealt{Totani:2008by,Maoz:2010qm,Maoz:2010dw}).  The physics behind this relation remains unclear, although such a distribution may naturally result from binary mergers (see, e.g., \citealt{Ruiter:2009dk}) or a single-degenerate scenario \citep{Hachisu:2008zw}.

\citet{Gallazzi:2005df} also derive $r$-band light-weighted galaxy ages, which vary from $\sim\! 10^9\,$yr at $10^9\,M_\odot$ to $\sim\! 10^{10}\,$yr at $10^{12}\,M_\odot$ (see their Fig.~8), using galactic models with an exponentially declining star formation history (SFH) from a time $t_{\rm form}$ with subsequent random bursts.  Ideally, one would have at hand the detailed history of star formation in every galaxy.  This is understandably difficult to achieve with any certainty.  Attempts have been made in this direction (e.g., \citealt{Brandt:2010jn,Maoz:2010dw}); however, using what amounts to an average over the galaxy population should be suitable for comparison with global rates.

If the \citet{Gallazzi:2005df} ages corresponded to a single-age stellar population at a given galactic mass, deriving the expected SN~Ia rate for a given DTD would be rather straightforward.  For example, using a DTD for each galaxy of the form
\begin{equation}
  \phi(\Delta t) = \phi_* \, \Delta t_{\rm Gyr}^{-\gamma}\,,
\label{dtd}
\end{equation}
with $t_{\rm Gyr}=t/$(1 Gyr), and assuming that the entire galactic stellar mass, $M_g$, arose at a single time, $t_g$, in Eq.~(\ref{eq:rate}) would lead to a galactic specific SN~Ia rate at time $t$ of
\begin{equation}
      \frac{\dot{N}_g(t)}{M_g} = \phi(t_{\rm Gyr}-t_{g,\,{\rm Gyr}}) = \frac{\phi_*}{(t_{\rm Gyr}-t_{g,\,{\rm Gyr}})^\gamma}\,.
\label{grate}
\end{equation}
This description is incomplete, though.  First, the SN~Ia rate at present reflects the galactic mass at the time of formation, as opposed to that measured today after stellar mass loss has occurred.  We correct for this using the results of \citet{Bruzual:2003tq} for a Chabrier IMF (as used in the SDSS galaxy sample) by including a term of the form $M(t_g)/M(t)$.

Additionally, the ages are more accurately galactic averages, so that an assumption of instantaneous formation at $t_g$ will not properly reflect the effect of a DTD.  To allow for a finite duration of star formation, we use a declining history of the form $e^{-t/\tau}$, with $\tau \!=\! 1$~Gyr, occurring since the time $t_g$ for each galaxy.  We further make use of the 16/84\% ranges in log~$t_g$ reported in \citet{Gallazzi:2005df} in order to weight the galaxy population with the DTD at fixed mass (rather than using only the median value).  These should alleviate the effect of average ages by giving greater weight to the low-age tail of their derived distribution and by allowing for a non-negligible rate of star formation today, particularly for lower-mass galaxies, that is in rough agreement with the specific star formation rates measured by \citet{Schiminovich:2007kw}.

This leaves the issue of the efficiency of converting a stellar population into SNe Ia (see, e.g., \citealt{Maoz:2007xw}).  As we will discuss in Section~\ref{section:v}, the DTD is also involved in shaping the cosmic SN~Ia history, with $\phi_*$ again setting the overall normalization.  Rather than attempting to incorporate theoretical models of the DTD (see, e.g., \citealt{Greggio:2010en,Meng:2011ba}), we use a DTD with a pure power law of the form $\Delta t^{-\gamma}$, with a lower cutoff $t_c$ to account for the minimum amount of time needed to produce CO WDs.  Evidence for delay times as short as $\lesssim\,$100~Myr has been reported from, e.g., the study of SN remnants in the Magellanic Clouds \citep{Badenes:2009bs,Maoz:2010pz}, and we simply use $t_c=50$~Myr (see also the discussion in Section~\ref{section:v}).

The results of using this approach are shown as the solid line in Fig.~\ref{Iarate}, where we have used a $\Delta t_{\rm Gyr}^{-1}$ DTD with $\phi_*=1.4\times 10^{-3}\,(10^{10}\,M_\odot)^{-1}\,$yr$^{-1}$.  A parametrization for this model is given in the Appendix.  We see that the saturation in age at high masses results in a plateau, which should be a rather robust feature due to the relatively small scatter in estimated ages around $\lesssim\! 10$~Gyr in this range, while the decrease in age at lower mass results in a rise in the SN~Ia rate.  Overall, this simplified model agrees rather well with the LOSS data.  Recently, \citet{Graur:2012na} followed the above prescription to compare with rates from a sample of SDSS SN~Ia hosts, finding a general agreement with our result.

\section{Incorporating Metallicity Dependence}
\label{section:iv}

Historically, studies have focused upon deriving the DTD without taking into account the possible effects of stellar metallicity on the SN~Ia rate in a galaxy.  If there is no such effect, then what we have done above would be sufficient.  As we next show, this assumption may prevent a determination of the actual DTD and hence its astrophysical origins.  We distinguish here between metallicity effects as primary (those involved in the rate of explosions) and secondary (those affecting the detailed properties of individual explosions; e.g., \citealt{Timmes:2003xx}), with our interest being in the former.

We now examine a plausible scenario for including a SN~Ia rate that varies with stellar metallicity.  We propose that this arises from the effect of metallicity on the white dwarf produced (near the end of this Section we discuss other possibilities).  In general, it is expected that, for the same initial stellar mass, the white dwarf from a star of lower metallicity should be more massive.  This may be due to decreased mass loss and/or opacity resulting in hotter burning over the lifetime of the star (e.g., \citealt{Umeda:1998ij,Willson:2000kb,Marigo:2007xq,Meng:2007ni}).  The simplest interpretation of this is that it should be easier to reach the requisite Chandrasekhar mass for explosion through the addition of mass via binary evolution or a double white dwarf merger.

To obtain a semi-quantitative estimate of the resulting change in the SN~Ia rate with metallicity, we must consider the effect of a varying WD mass over the range of metallicities for the galaxies in the LOSS sample.  One may hope for guidance from the initial-final WD mass relation determined from young star clusters. However, the clusters in which detailed studies are possible are nearby and formed recently, which necessarily limits them to single, $\sim\,$solar metallicity stars (e.g., \citealt{Kalirai:2007tq,Williams:2008ms}).  We utilize the theoretical results of \citet{Umeda:1998ij} in combination with the empirical metallicity estimates in \citet{Gallazzi:2005df}.  In decreasing the initial stellar metallicity from $Z \!=\! 0.03$ to 0.004, roughly the range spanned in Fig.~\ref{Iarate}, \citet{Umeda:1998ij} determined that an additional $\sim\! 0.05 \!-\! 0.15\,M_\odot$ is added to the CO remnant at fixed initial mass (see their Fig.~6).

Fig.~8 of \citet{Umeda:1998ij} displays the relative number of SN~Ia progenitors obtained from their stellar evolution model as a function of metallicity after integrating over a Salpeter IMF from a lower initial stellar mass (which varies with $Z$) corresponding to fixed final WD mass to an upper mass at which point ONeMg WDs were expected to be produced (varying with $Z$ from $\sim\! 7-8.5\, M_\odot$).  Using a threshold WD mass of $0.85\,M_\odot$ yields a dependence on the rate with metallicity that can be approximately parametrized as
\begin{equation}
     N_{\rm Ia}(Z) \propto \left(Z+0.003\right)^{-0.5}
\label{umfit}
\end{equation}
over the range $Z \!=\! 0.004 \!-\! 0.03$.  Using a lower threshold mass of $0.7\,M_\odot$ yields a slightly weaker dependence, due to the larger mass range, of approximately
\begin{equation}
     N_{\rm Ia}(Z) \propto \left(Z+0.0015\right)^{-0.3}.
\label{umfit2}
\end{equation}
To derive galactic rates, we again use a SN~Ia rate for each galaxy $\propto\! \Delta t^{-\gamma}$ and scale directly to the \citet{Umeda:1998ij} results, normalizing these relations to unity at $Z \!=\! 0.025$, the metallicity of a characteristic $\sim\! 10^{11}\,M_\odot$ galaxy in the LOSS sample.  Assuming $Z_g$ and $t_g$ to be separable, we use the 16/84\% ranges in log~$Z_g$ from \citet{Gallazzi:2005df} to again weight the galaxy distribution at fixed mass and introduce an overall term to account for the effect of metallicity in Eq.~(\ref{eq:rate}), either $f_{0.85}(Z)$ or $f_{0.70}(Z)$.

The specific SN~Ia rates resulting from using the two metallicity scalings are shown in Fig.~\ref{Iarate}.  For the $f_{0.85}$ model, $\phi_* = 1.1\times 10^{-3}\,$$(10^{10}\,M_\odot)^{-1}\,$yr$^{-1}$ with $\Delta t_{\rm Gyr}^{-0.8}$ (dashed line), while the $f_{0.70}$ model has $\phi_* = 1.3\times 10^{-3}\,(10^{10}\,M_\odot)^{-1}\,$yr$^{-1}$ and $\Delta t_{\rm Gyr}^{-0.9}$ (dotted line).  After accounting for the weaker effects of the DTDs used, the rate does indeed rise more steeply at lower galactic masses than by taking into account age alone.  This can be interpreted as a relative change in efficiency, an effect at the factor of $\sim\,$2 level over the mass range of Fig.~\ref{Iarate} for the $f_{0.85}$ model and slightly less for the $f_{0.70}$ case.

Because galaxy mass is strongly correlated with both age and metallicity, it is inevitable that the models are relatively degenerate and that inferences about the DTD from galaxy populations may err without accounting for metallicity.  Our simplified treatment of galactic star formation histories may somewhat underpredict SN~Ia rates at intermediate masses.  This may be refined through more detailed modeling, although, given the uncertainties in our inputs, we will not attempt to do so here.  It is encouraging that such broad agreement with data is already seen using quite general assumptions.

We note here that the normalization of these models can be scaled up or down, although this will directly affect the normalization of the expected cosmic SN~Ia rates through the DTD, as we will discuss in the following Section.  Note also that care should be taken in comparing these results, which examine the galaxy population as a whole, to those that distinguish between ``passive'' and ``star-forming'' galaxies (e.g., \citealt{Sullivan:2006ah}).

The above is essentially based on an assumption of a single-degenerate scenario.  While the full effects of metallicity on a double-degenerate scenario are likely more complicated, if we consider binaries with a uniform mass ratio distribution, the number that will exceed the Chandrasekhar limit depends on metallicity as
\begin{equation}
     N_{\rm Ia} \propto Z^{(x-1)\,b/a} \sim Z^{-0.4}\,,
\label{binz}
\end{equation}
where $x=2.35$ is the slope of the IMF, $a=0.5$ approximates the slope of the white dwarf initial-final mass relation of \citet{Kalirai:2007tq}, and $b \simeq -0.08$ is the dependence of the final mass on metallicity, estimated from \citet{Umeda:1998ij}.  The magnitude of the effect is very similar to the case already considered, so we do not repeat the calculations.  This model does not include any effect of the higher implied masses on the rate of binary evolution or possible effects in triple systems based on the Kozai mechanism \citep{Thompson:2010dp}.

\begin{figure}[b!]
\includegraphics[width=3.35in,clip=true]{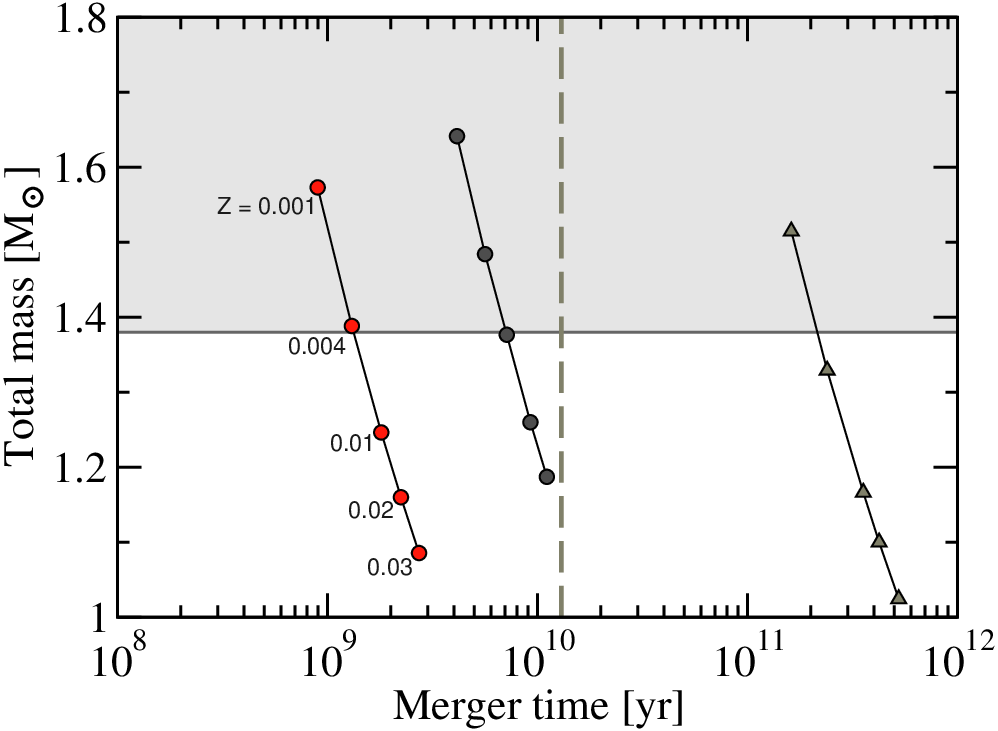}
\caption{The total masses of known WD-WD systems along with their calculated merger times due to gravitational wave losses.  Shown are the three systems from the collection in \citet{Nelemans:2005qb} in which both WDs have masses exceeding 0.5~$M_\odot$: two with firm masses ({\it circles}) and one with only a lower limit on the mass of the secondary ({\it triangles}).  Assuming these to have all resulted from stars with $Z\sim0.02$, we show the ``expected'' total masses for a range of metallicities (as labeled) using the final masses derived in \citet{Umeda:1998ij}.  The resulting merger times assume initial orbital separations as presently inferred for each.
\label{masses}}
\end{figure}

To illustrate the above effect, we begin with the three known double WD binaries in which each component has a mass of at least $0.5\,M_\odot$ from \citet{Nelemans:2005qb}.  These have primary/secondary masses of $0.71+0.55\, M_\odot$, $0.58+0.58\, M_\odot$, and $0.51\,+ $ $>$$0.59\, M_\odot$, the last being a single-lined system with only a lower limit for the secondary.  We assume that each WD arose from a solar-metallicity star and map from the WD masses to the initial stellar masses using the results of \citet{Umeda:1998ij}, mapping then to the WD masses calculated for these stellar masses at other metallicities.  The $0.5\,M_\odot$ cut allows a straightforward translation without regards to systems with low-mass He WDs, etc.  Keeping the initial binary separations fixed, we calculate the merger time due to gravitational wave losses for each system (see, e.g., \citealt{Thompson:2010dp}).  Fig.~\ref{masses} displays the effect on these systems using this prescription, where it is seen that the total masses of all three systems would have been pushed beyond the Chandrasekhar mass limit and the merger time would have been significantly reduced at lower metallicities.

There is hope for new tests to reduce the uncertainty in the overall effect of metallicity.  For example, in observations of SN~Ia host galaxies we would expect the hosts of SN~Ia to be slightly less metal-rich than the galaxy population as a whole for fixed galaxy mass.  This effect would not be as marked as in the case of gamma-ray bursts (see \citealt{Stanek:2006gc}), since no hard upper metallicity threshold prohibiting the production of a SN~Ia progenitor system is known to exist.  An exception to this may be found at very high metallicity, as evidenced by the abundance of He rather than CO WDs in the metal-rich cluster NGC~6791 (see \citealt{Kilic:2007yk}) -- very massive, metal-rich galaxies may show an additional deficit of SNe~Ia beyond that of our simple model.  This may even be evident in the data at the high-mass end of Fig.~\ref{Iarate}, although it is difficult to draw a strong conclusion at present.

We note that the model that we have used only results in a rather modest rate change with metallicity.  It does not attempt to account for changes in the remnant mass that occur during the AGB phase of an isolated star (see, e.g., \citealt{Vassiliadis1993,Bird:2010yz,Renedo:2010vb}), which could result in a larger metallicity effect.  Since both SN~Ia scenarios require binary evolution at some step in their evolution, this model should be adequate in this respect.  We also have not attempted to vary the binary fraction with stellar mass or metallicity.  This is not yet well understood either theoretically or empirically (see, e.g., \citealt{Mazeh:2006vh}), particularly in the mass range of the progenitors of SNe~Ia.

In addition to a possible diminishment of stellar winds at low $Z$ \citep{Kobayashi:1998ii}, another effect that may work in the opposite direction, particularly for single-degenerate scenarios, is the increased compactness of lower metallicity stars, which can limit interactions \citep{deMink:2007if} and work towards a rate suppression.  As a further complication, though, this aspect might also allow more such stars to reach core helium burning and form a CO white dwarf before mass transfer does occur \citep{deMink:2007gh}.  The net effect of varying stellar radii along with the number and masses of resulting white dwarfs remains to be determined and should be considered in more detail at the level of binary evolution and in population synthesis modeling.  All this suggests that substantial room for improvement exists on both the theory and observing fronts and an initial goal should be to determine the overall sign of the influence of metallicity on SN Ia rates.

\section{The Cosmic Type Ia Supernova Rate}
\label{section:v}

\begin{figure}[t!]
\includegraphics[width=3.38in,clip=true]{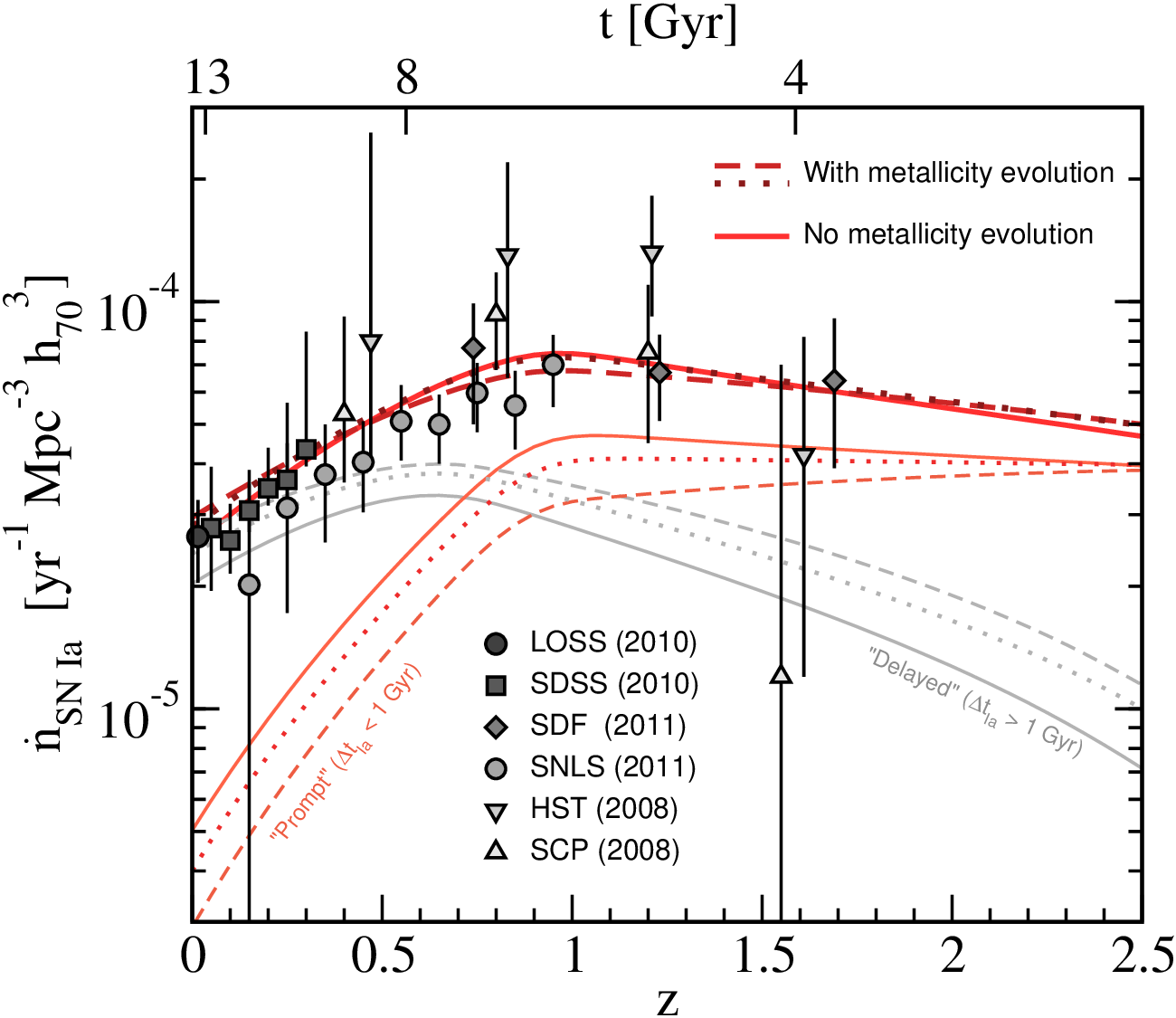}
\caption{The cosmic rate of Type Ia supernovae.  Shown are recent measurements from LOSS \citep{Li:2010iii}, SDSS \citep{Dilday:2010qk}, SCP \citep{Kuznetsova:2007ew}, HST \citep{Dahlen:2008uj}, SNLS \citep{Gonzalez} and Subaru Deep Field \citep{Graur}.  A model assuming only a fixed $\Delta t^{-1}$ delay time distribution and the cosmic SFR from \citet{Kistler09} ({\it thick solid line}) can be compared to our models incorporating metallicity dependence (see text), which use either a $\Delta t^{-0.9}$ DTD ({\it thick dotted line}) or $\Delta t^{-0.8}$ DTD ({\it thick dashed line}).  The components of these models with delays from stellar birth to explosion of less than 1~Gyr (``prompt'') and greater than 1~Gyr (``delayed'') are also shown ({\it thin lines}; as labelled).\\
\label{Iahist}}
\end{figure}

We next examine the expectations for the cosmic rate of SNe~Ia by again first considering a case without explicit metallicity dependence.  We proceed by returning to Eq.~(\ref{eq:rate}) with the comoving star formation rate density $\dot{\rho}_*(z)$ inferred up to $z \sim 8$, using the \citet{Yuksel:2008cu} parametrization of the SFH,
\begin{eqnarray}
      \dot{\rho}_*(z)
      & = &  \dot\rho_0 \left[(1 \!+\! z)^{{a}{\eta} } + \left(\frac{1 \!+\! z}{B}\right)^{{b}{\eta}} + \left(\frac{1 \!+\! z}{C}\right)^{{c}{\eta} } \, \right]^{1/\eta}
\label{fit}
\end{eqnarray}
where $a \!=\! 3.4$,  $b \!=\! -0.3$, and $c \!=\! -2$, with breaks at $z_1\!=\! 1$ and $z_2 \!=\! 4$ corresponding to $B \!=\! (1 \!+\! z_1)^{1-a/b} \!\simeq\! 5100$ and $C \!=\! (1 \!+\! z_1)^{(b-a)/c} (1 \!+\! z_2)^{1-b/c} \!\simeq\! 14$, which reflect the updated high-$z$ data from \citet{Kistler09}, and we use $\eta \!\simeq\! -10$ to smooth the transitions.  The normalization is $\dot{\rho}_{0} \!=\! 0.014 \,M_\odot$~yr$^{-1}$~Mpc$^{-3}$, which we have scaled down by a factor of 0.7 from the Salpeter IMF normalization of \citet{Hopkins:2006bw} to be in better agreement with the galactic mass estimates in the previous Sections.  In Eq.~(\ref{eq:rate}), we set $t_0 \!\sim\! 0.4$~Gyr corresponding to $z \!\approx\! 10$.\footnote{We use $\Omega_{\rm m} \!=\! 0.3$, $\Omega_\Lambda \!=\! 0.7$, and $H_0 \!=\! 70$~km/s/Mpc where needed, e.g., in converting $z\leftrightarrow t$ and in rescaling the data in Fig.~\ref{Iahist} using a common value of $H_0$.}

Since we must consider the shortest possible delay times in constructing the cosmic rate from the SFH, we again assume a power law DTD with a cutoff at $t_c \!=\! 50$~Myr.  We note that the results of \citet{Umeda:1998ij} (see also \citealt{Siess2007,Meng:2007ni}) suggest a maximum CO WD mass of $\sim\! 1.1\,M_\odot$ that is nearly independent of metallicity.  Since the effect of decreasing the metallicity is similar to increasing the stellar mass, we take this cutoff to be independent of $Z$ (and thus $z$) since the lifetimes of the stars giving the most massive CO WDs should be similar.

We use the $\Delta t_{\rm Gyr}^{-1}$ DTD and $\phi_*$ obtained by comparison to the LOSS data in Section~\ref{section:iii}, so that the resulting SN~Ia rate history parameters are fixed, leading to the evolution shown in Fig.~\ref{Iahist} (thick solid line).  To compare with prior results (e.g., \citealt{Mannucci:2005xh,Scannapieco:2005uh,Sullivan:2006ah}), this history is broken down into the components with delay less than 1~Gyr (thin solid lines labelled as ``prompt'') and greater than 1~Gyr (labelled as ``delayed'').  We see that the ``prompt'' component is subdominant at $z \!=\! 0$, in agreement with the rates in Fig.~\ref{Iarate}.  Altering either the form of the DTD or $t_c$ can make the ``prompt'' component relatively more or less important (see, e.g., \citealt{Horiuchi} for related discussion); however, this would in turn affect the specific SN~Ia rate models in Sections~\ref{section:iii} and \ref{section:iv}.

Since the universe as whole had a lower metallicity in the past, a relative enhancement should also be effected in the cosmic SN~Ia rate.  As existing rate measurements average over the entirety of the galaxy population, this effect should not be dramatic at the present epoch, but, as for the specific rate, can be important in deriving the DTD.  At low $z$, the gas-phase metallicity is typically higher than that of the stellar population \citep{Gallazzi:2005df}.  The relation between galaxy mass and gas-phase metallicity is well determined at low $z$ \citep{Tremonti:2004et} and has been measured to evolve at higher redshifts, so that the typical metallicity decreases by $\sim\! 0.15$~dex per $z$ up to at least $z \!\approx\! 2$ (e.g., \citealt{Kewley,Savaglio:2005hi,Erb:2006qy,Maiolino:2008gh}).  We use
\begin{equation}
      Z(z) = 0.03 \times 10^{-0.15\,z}
\label{Zztop}
\end{equation}
to account for stars forming from gas that is increasingly metal poor at higher $z$, with a resulting change in rate arising through either the relation approximated by Eq.~(\ref{umfit}) for the $\dot{n}_{0.85}$ model or Eq.~(\ref{umfit2}) for the $\dot{n}_{0.70}$ model, again normalizing each to unity at $Z \!=\! 0.025$ to be consistent with our specific rate models.  We also use the same values of $\phi_*$ and DTD slopes as in the corresponding specific rate models.

Fig.~\ref{Iahist} shows the resulting cosmic rates for the $\dot{n}_{0.85}$ (with $\Delta t_{\rm Gyr}^{-0.8}$; thick dashed line) and $\dot{n}_{0.70}$ ($\Delta t_{\rm Gyr}^{-0.9}$; thick dotted line) models.  Both models yield similar histories as the metallicity-independent case, with parametrizations for all three included in the Appendix.  This is due to the relative increase of the rates with $z$ as compared to models with the same DTD without a metallicity enhancement.  This is similar in spirit, but less dramatic, than the relative evolution likely due to stellar metallicity seen in the cosmic GRB rate (e.g., \citealt{Kistler}).  Both models are also broken down by delay time in Fig.~\ref{Iahist} (thin dotted, dashed lines), which illustrates the underlying effect of altering the DTD.

As discussed for the specific SN~Ia rate, there is again a degeneracy between altering the DTD and including the effect of metallicity, although not quite as strong.  That the metallicity effect works in the same direction as decreasing the index in the DTD in both cases, as seen in Fig.~\ref{Iarate} and Fig.~\ref{Iahist}, is something of a coincidence, owing to the fact that galaxy ages and metallicities both decrease with decreasing mass and the cosmic SFR rises with increasing $z$.  This didn't have to be the case, though.  We thus reiterate that an estimate of one component must account for the other until this degeneracy is broken.  It is possible to perform a more elaborate study by varying all the parameters involved (see, e.g., \citealt{Horiuchi} and \citealt{Graur} for the metallicity-independent DTD); however, the qualitative effects of the models that we have considered are already sufficiently evident.

\section{Discussion and Conclusions}
\label{section:vi}

The rate of Type Ia supernovae should be affected at some level by the effects of metallicity on stellar evolution.  There may be various complications involved, such as the largely unresolved effects of binary evolution, but our simple model for the effects of metallicity should be broadly relevant.  There has been significant effort devoted to investigating Type~Ia SN properties as a function of metallicity (e.g., \citealt{Hamuy:2000ya,Gallagher:2008zi,Howell:2008jv,Neill:2009tr,Sullivan:2010mg,Konishi:2011ct}).  Since the properties of SN~Ia have been seen to depend on metallicity, why not the rate?

The simple models that we have considered explain fairly well both the specific SN~Ia rates measured in nearby galaxies by LOSS and the observed normalization and evolution of the cosmic SN~Ia rate.  An enhanced rate due to more massive white dwarf remnants is in contrast to other effects that may be important during binary interactions, such as a decrease of stellar winds or radius.  Future models in this area can utilize the framework presented here as a basis for comparison with data.  Attempts to introduce the various effects discussed into population synthesis models are welcome.  Since the full problem of solving the complete chain of events of binary evolution leading to a SN~Ia is a very difficult problem, a reasonable first step would be to determine empirically whether there is a net enhancement or suppression.

A low-$Z$ enhancement leads to an expectation of a relatively higher SN~Ia rate in the outer regions of galaxies, for a given stellar population age and total mass, due to the lower average metallicity.  It is thus important to take into account not just the integrated metallicity of the galaxy, but the value at the birthplace of the progenitor.  Such a bias may already be seen in the number of cases in which a SN~Ia occurred in the outer halo of a star-forming galaxy \citep{Prieto,Khan:2010kj}.  Using the location of the explosion as a proxy in such a differential study would be cleanest performed by considering ``prompt'' SNe~Ia and/or small, more metallicity-homogeneous galaxies.

Further progress can certainly be made with data that can thus suitably break the degeneracy between decreasing metallicity and decreasing age.  Evidence in this direction has been found in a comparison of SN~Ia host galaxies in SDSS by \citet{Cooper:2009vk}, who found that SNe~Ia in blue, star-forming hosts have a preference for lower-density environments, which they interpreted as being the effect of lower gas-phase metallicities.  Non-targeted SN searches are useful in this regard, such as ROTSE-IIIb, which found an excess of dwarf hosts in their SN~Ia sample \citep{Quimby:2012va}.

New efforts to discover ever higher-redshift SNe~Ia are also necessarily probing a regime where the intrinsic metallicity is lower.  Such surveys have recently uncovered two SNe~Ia at $z \!\simeq\! 1.55$ with very-low $Z$ hosts, one with 12 + log(O/H) = 8.12$^{+0.09}_{-0.10}$ \citep{Frederiksen:2012px} and another with the rather low 12 + log(O/H) $<$ 8.0 \citep{Frederiksen:2012qd}.  Taking these as upper limits on the SN progenitor stellar $Z$ pressures models with a metallicity floor.  The model of \citet{Kobayashi:1998ii} in particular, with a minimum metallicity of $\sim\! 0.1\,Z_\odot$, predicts a negligible number of SNe~Ia at $z \!\gtrsim\! 1$.  In contrast, the continued observation of high-$z$ supernovae, such as the recent $z \!=\! 1.914$ event \citep{Jones:2013dta}, is expected in the picture we advance.  Since the evolution with $z$ of the $M\!-\!Z$ relation is seen to proceed more rapidly at low masses \citep{Zahid:2013pla}, we would thus anticipate relatively more discoveries of such prompt  SNe in low-mass hosts, which as discussed above are favored for relative rate analyses.

Additionally, observations of Milky Way dwarf spheroidal galaxies have revealed decreasing values of [$\alpha$/Fe] with increasing [Fe/H], indicating the influence of Type~Ia supernovae down to metallicities of [Fe/H] $\approx -2.5$ \citep{Kirby:2010dc}.  Since model fits to these measurements are naturally sensitive to the SN~Ia rate over a range of metallicities, we urge exploration of the implications of an increased rate at low~$Z$, including super-Chandrasekhar mergers, on galactic chemical evolution.

As previously mentioned, the results of \citet{Umeda:1998ij} indicate that the maximum CO WD mass remains close to $\sim\! 1.1\,M_\odot$ over a wide range of metallicities.  If this is true, and binary evolution effects are neglected, then we would expect the relative rates of super-Chandrasekhar SNe~Ia arising from mergers to increase with lower metallicity in proportion to the normal SNe~Ia due to the power law form of the IMF (note that instabilities prohibit the necessary growth of even rapidly-rotating single WDs; \citealt{Piro:2008pr}).  However, recent observations of host galaxies may indicate an even stronger preference for low-metallicity hosts for super-Chandrasekhar SNe~Ia (e.g., \citealt{Taubenberger:2010qv,Childress:2011hg}).  Moreover, the maximum mass resulting from a merger under these assumptions is $\sim\! 2.2\,M_\odot$, below the $2.4 \!\pm\! 0.2\,M_\odot$ total mass inferred from SN~2007if \citep{Scalzo:2010xd}.

An explanation for both effects may arise from binary evolution.  To achieve a higher total merger mass, without resorting to an ONeMg WD, at least one WD should gain mass while maintaining a CO composition.  If the effect of inhibiting single degenerate SN~Ia production at $Z \lesssim 0.1\,Z_\odot$ \citep{Kobayashi:1998ii,Kobayashi:2008se} does hold, then the primary WD may instead be pushed close to, but not above, the threshold for explosion, so that the rate of massive mergers is further enhanced.  The end state of the secondary resulting in a massive CO WD could then lead to a merger with total mass upwards of $\sim 2.5\,M_\odot$.

If, contrary to \citet{Umeda:1998ij}, the CO WD mass limit actually increases modestly for lower metallicities, then the rate of such extreme super-Chandrasekhar mergers rises dramatically at lower metallicities without a need to turn to binary evolution for a solution.  This is due to the presence of the threshold in reaching the requisite total merger mass, which would lead to a large relative difference between low/high-$Z$ galaxies.  The stars giving rise to these massive white dwarfs would also evolve more rapidly and could thus lead to ``prompter'' explosions in low-$Z$ environments.  Whether this scenario occurs is a question left for stellar evolutionary modeling and observations of host galaxies.  We note that elevated rates of other transients involving a white dwarf and dependent upon the mass (e.g., \citealt{Thompson:2009km}) could also be expected.

In the category of interesting, but more anecdotal, evidence that low metallicity might be of significance for Type Ia supernovae, \citet{Tovmassian:2010uq} recently presented strong evidence that SBS1150+599A, a close binary star inside a metal-poor, Galactic halo planetary nebula PN G135.9+55.9 consists of two white dwarfs that will merge within a Hubble time. The estimated total mass of the binary is very close to the Chandrasekhar limit, making it a likely SN~Ia progenitor.

It is also interesting to note that the normalized rate of planetary nebulae in elliptical galaxies \citep{Buzzoni:2006ei} shows a very similar trend with metallicity to that discussed in our Fig.~1. Indeed, their Figs.~11 and 12 show about 10 times fewer planetary nebulae per unit luminosity in metal-rich, massive ellipticals compared to metal-poor, low-mass ellipticals.  The mapping between PN production and SN~Ia explosion is of course uncertain; however, both involve the production of a white dwarf, and \citet{Buzzoni:2006ei} attribute finding fewer PNe in more metal rich ellipticals to a dependence of the initial-to-final mass relation on metallicity.

Substantial observational progress has been made in the study of SNe~Ia in the last decade and new data can be expected to better determine the extent to which metallicity affects the SN~Ia rate.  As discussed above, the possible directions include detailed measurements of rates within galaxies to examine the $Z$ dimension.  Improved measurements of the cosmic SN~Ia rate, in combination with independent determinations of the DTD at fixed $Z$, can examine whether the rate is larger than otherwise expected.  Along with these, if the intrinsic properties of SNe Ia vary with metallicity, evolution in the Type Ia luminosity function can complicate cosmological determinations \citep{Riess(2006)}, which in the case presented here would be more pronounced as the lower-$Z$ component becomes further enhanced at higher redshifts.  This places added emphasis on the importance of determining the net effect of metallicity on the SN Ia rate, and if a null result is eventually established, how the various effects discussed here could conspire in such a way.


\appendix
\label{section:app}

The models that we have discussed are the result of combining several unrelated functions and thus are not necessarily amenable to convenient parametrization.  Nonetheless, we find that a sigmoid function provides an adequate fit to our metallicity-independent model of the specific SN~Ia rate, $\zeta_{\rm Ia}$, with
\begin{equation}
      \frac{\zeta_{\rm Ia}(M)}{(10^{10}\,M_\odot)^{-1}\,{\rm yr}^{-1}} = \alpha \, \left[1+\exp\left(\frac{\log (M/M_\odot)-M_*}{\omega}\right)\right]^{-1}+\beta\,,
\label{zfit}
\end{equation}
where $\alpha=5\times10^{-3}$, $\beta=4.2\times10^{-4}$, $M_*=10$, and $\omega=0.33$ agrees with the model to within $<10\%$ over the mass range displayed in Fig.~\ref{Iarate}.  The metallicity-dependent models can be fit with similar parameters.

Using the smoothly-broken piecewise form of Eq.~(\ref{fit}), with $\dot{\rho}_0$ replaced by $\dot{n}_0$, our cosmic rate models can be fit to within a few percent over the range $z=0-4$.  The parameters used for the metallicity-independent model and the metallicity-dependent $\dot{n}_{0.85}$ and $\dot{n}_{0.70}$ models are given below in Table~\ref{tab:params}.  All three use $\eta \simeq -10$ to smooth the transitions.

%
\begin{table}[h]
\caption{Parameters used in the fits of our three cosmic SN~Ia rate scenarios.}
\label{tab:params}
\begin{ruledtabular}
\begin{tabular}{lcccccccc}
Model         & $\dot{n}_0$ [yr$^{-1}$~Mpc$^{-3}$]                & $a$          & $b$        & $c$	            & $z_1$   &   $z_2$  &  $B$    &  $C$  \\ \hline
$Z$-free     & $2.5\times10^{-5}$    & $1.8$       & $-0.8$   & $-2.3$	   & $0.9$    &   $2.9$   &  $8.1$ &  $5.0$\\
$\dot{n}_{0.85}$  & $2.9\times10^{-5}$    & $1.4$       & $-0.5$   & $-2.0$	   & $0.9$    &   $2.9$   &  $11.5$ &  $5.1$ \\
$\dot{n}_{0.70}$  & $2.8\times10^{-5}$    & $1.6$       & $-0.7$   & $-2.0$	   & $0.9$    &   $3.0$   &  $8.2$ &  $5.2$ \\
\end{tabular}
\end{ruledtabular}
\end{table}
%


\acknowledgments

We thank John Beacom, Jonathan Bird, Shunsaku Horiuchi, Rubab Khan, Marc Pinsonneault, and Hasan Yuksel for helpful discussions, Weidong Li for providing us with the data in Fig.~1, and the Referee for helpful comments.
M.D.K.\ acknowledges support provided by NASA through the Einstein Fellowship Program, grant PF0-110074.
K.Z.S., C.S.K.\ and T.A.T.\ are supported in part by NSF grant AST-0908816.
J.L.P.\ acknowledges support from NASA through Hubble Fellowship grant HF-51261.01-A awarded by the STScI, which is operated by AURA, Inc. for NASA, under contract NAS 5-26555.
T.A.T.\ is supported in part by an Alfred P.~Sloan Foundation Fellowship.


\end{document}